\documentstyle[aps,prl,amssymb,floats,twocolumn,psfig]{revtex}

\begin{document}

\wideabs{

\title{Wideband dual sphere detector of gravitational waves}

\author{M.\ Cerdonio$^1$, L.\ Conti$^1$\footnote{Corresponding author: conti@lnl.infn.it},
J.A.\ Lobo$^2$, A.\ Ortolan$^3$, L.\ Taffarello$^4$
and J.P.\ Zendri$^4$}

\address{$^1$INFN\ Padova Section and Department of Physics, University of
Padova, via Marzolo 8, I-35100 Padova, Italy}
\address{$^2$Departament de F\'\i sica Fonamental, Universitat de Barcelona,
Diagonal 647, 08028 Barcelona, Spain}
\address{$^3$INFN, Laboratori Nazionali di Legnaro, via Romea 4, I-35020
Legnaro, Padova, Italy}
\address{$^4$INFN\ Padova Section, via Marzolo 8, I-35100 Padova, Italy}

%
%
%
%

\date{$\,^*$Corresponding author: conti@lnl.infn.it}

\maketitle

\begin{abstract}
We present the concept of a sensitive {\itshape{and}} broadband resonant
mass gravitational wave detector. A massive sphere is suspended
inside a second hollow one. Short, high-finesse Fabry-Perot optical
cavities read out the differential displacements of the two spheres
as their quadrupole modes are excited. At cryogenic temperatures one
approaches the Standard Quantum Limit for broadband operation with
reasonable choices for the cavity finesses and the intracavity light
power. A molybdenum detector of overall size of 2 m, would reach spectral
strain sensitivities of $2\cdot 10^{-23}$~Hz$^{-1/2}$ between 1000~Hz and 3000~Hz.
\end{abstract}
\

}	


Resonant mass detectors of gravitational waves (GW) are commonly indicated
as narrowband devices. In currently operating cylindrical bar
detectors~\cite{bars}, all equipped with resonant transducers, the
bandwidth, even in the limit of quantum limited performance of the
final displacement readout, would not open up for more than about the
frequency interval between the resulting two mechanical modes of resonance,
currently about 30~Hz.

This stems however from the noise performance of the employed readout systems. A general analysis of
the problem, valid for any {\itshape linear} detector, has been given in ref.~\cite{Price}. Secondary {\itshape resonant} masses are linked to
the main resonant mass to efficiently couple the signal amplitude to the
final readout, but then the bandwidth is limited to a fraction of the main resonator
frequency. Such a coupling is poorer the smaller the total number $n\/$ of
resonators, and correspondingly the bandwidth decreases with $n\/$. To open the bandwidth one would have to use multimode systems~\cite{paik,Richard} with $n\,\geq 3$, but to date only two-(mechanical) mode systems have worked their way into
operating detectors, giving a fractional bandwidth $\Delta f/f \ll 0.1$.

However, if a {\itshape single} mechanical
resonator were driven only by its thermal noise and by a signal force the signal to noise ratio (SNR) would be {\itshape independent}
of frequency, and thus the band would open up {\it provided} enough signal amplitude can be coupled to the final readout. The possibilities offered nowadays by optomechanical systems are such that the interplay between the back-action of the radiation pressure and the photon counting noise in a high finesse, high power Fabry-Perot cavity would allow enough signal coupling to get broadband operation at the SQL~\cite{Req}.

We have been attracted by the possibilities offered by optical readout
systems, as vigorously developed for interferometric GW detectors, and
more recently applied in connection with cryogenic bar GW
detectors~\cite{Richard92,Conti}. We take a Fabry-Perot optical cavity as
the motion sensor. In a system under development~\cite{Conti} the
length of the sensor cavity is compared to that of a second cavity,
separately kept, which acts as reference. We do not take into account
here the noise introduced by the reference cavity, assuming for simplicity
that it is negligible. With a sensor cavity length of the order of
centimeters there is no loss of signal strength for finesses $F\/$ as high
as the highest attainable with current technology, $F = 10^6$, for GW
in the kHz range. So we have considerable freedom
to vary the finesse and the light power $P$ incident on the cavity,
in search for optimal conditions at a chosen frequency, which do not demand unreasonable
values for these parameters. 

Let us then turn to the primary mechanical resonator, whose motion is
directly related to the incoming GW. We take into consideration
both solid and hollow spheres as resonant systems of interest. They are
very attractive for a number of reasons, and in fact they have received 
significant attention in the literature of the last few
years~\cite{jm,zu,lobo,vega}. Spherical detectors are omnidirectional,
have a more efficient coupling to the GW field relative to cylindrical
bars, both in the first and in the second quadrupole mode, and enable a
deconvolution of the GW signal if they are equipped with five (or more)
suitable motion sensors~\cite{jm,zu,lm}. For instance, a chirp signal from
a merging compact binary can be fully deconvolved with a spherical
detector~\cite{cocciaf}. Two spheres would make up for a complete
observatory, in which all parameters characterizing the incoming wave (e.g. velocity, 
direction, polarization) can be resolved~\cite{cerdoam1,clo}
-see also~\cite{massimoprl}. Located close to an interferometric
detector, a spherical detector could be used for searches of stochastic
background~\cite{Vitale}. Such capabilities make of the spherical detector
a conceptually unique device.

However, the sensitive spherical detectors proposed in the past suffer
from the above discussed bandwidth limitation. As an example, a hollow
sphere of CuAl$^{10\%}$, 4~m in diameter and 0.3~m thick,
cooled at sub-Kelvin temperatures and equipped with {\itshape{resonant}}
transducers and a quantum limited readout, gives a spectral strain noise
as low as $6 \cdot 10^{-24}$~Hz$^{-1/2}$~\cite{vega}, but only in two bands
of 35~Hz and 135~Hz, respectively around the first and second quadrupole
resonances at 350~Hz and at 1350~Hz.

\begin{figure}[!htbp]
\centerline{\psfig{figure=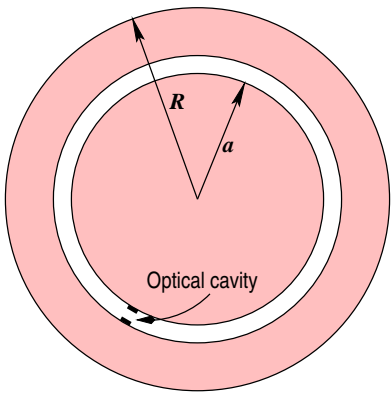,width=5.5cm}}
\caption{A dual sphere GW detector with Fabry-Perot
cavities as motion sensors.}
\label{fig1}
\end{figure}

Let us then consider a spherical detector with {\itshape{non}} resonant
optical readout. We need to integrate the two mirrors of each
Fabry-Perot sensing cavity in two separate systems, which must be cold,
massive and of high mechanical $Q\/$ factor, otherwise the thermal noise would be unacceptably large.
We are thus led to the concept of a GW detector based on a massive
{\itshape dual sphere} system of resonators: a hollow sphere which
encloses a smaller solid sphere, see Fig.~\ref{fig1}. Motion sensors in this
system will be optical Fabry-Perot cavities formed by mirrors coated face
to face to the inner surface of the hollow sphere and to the solid sphere,
in either a PHC~\cite{lm} or a TIGA~\cite{jm} layout.

The main sources of noise are: thermal noise in the large detector masses, back-action noise
introduced by the radiation pressure, and photon counting noise.
Given the optical figures, the evaluation of all three contributions for
our design is straightforward, as the spectrum of the resonant frequencies
of the two spheres is known~\cite{lobo,vega} once their material(s) and
dimensions are fixed.

Assuming the same material is used for both spheres, and that the inner
one of radius $a\/$ fills up almost completely the interior of the other
(external radius $R$, internal radius $\gtrsim a$), the first quadrupole
resonance of the outer hollow sphere is at the lowest frequency, while that
of the inner solid sphere is 2-3 times higher. The frequency region in
between is of particular interest: the GW signal drives the hollow
sphere above resonance and the solid sphere below resonance. The
responses of the two resonators are then out of phase by $\pi$ radians
and therefore the differential motion, read by the optical sensors,
results in a signal enhancement. In this region only a small number of
non-quadrupole resonances occur, which are not GW active. The pattern
repeats for the two second quadrupole modes at higher frequency and so on.
At a few specific frequencies above the first quadrupole resonance of the
solid sphere, under the combined effect of the response to GW of all the
quadrupole modes, their responses subtract and the sensitivity is reduced
and eventually lost in a few narrow bands. In this higher frequency region,
in addition to such loss of response, several resonances from the
GW-inactive modes appear. While the spectral sensitivity would be still
of some interest, we prefer for brevity to not discuss it here.

Let us assume that we only sense {\itshape radial} displacements and that
the spherical symmetry of the resonators is not broken; suspending a solid spherical resonator has shown to alter only marginally the spectrum of resonances~\cite{jm}. Using the notation
of references~\cite{lobo} and~\cite{vega} the response to a GW of the solid
sphere at its surface and of the hollow at its {\itshape inner} surface
(i.e. at the radius $a\/$) are respectively given by expressions of the
type

\begin{equation} 
u(\omega) = -\frac{1}{2} \sum_{n=0}^\infty b_n A_{n2}(a)\,
\omega^2 \tilde{h}(\omega)\,L_{n2}(\omega)      \ ,
\end{equation} 
where $A_{n2} (a)$ are radial function coefficients, $b_n$ are the
coefficients in the orthogonal expansion of the response function of each
sphere, $L_{n2} (\omega)$ is the Lorentzian curve associated to the mode
$\{n2\}$, the $n\/$-th quadrupole harmonic, and
$\tilde{h}(\omega)\equiv\tilde{h}_{ij}(\omega)\,n_i n_j$ is the Fourier
amplitude of the GW strain at the sensing point direction, defined by the
unit radial vector {\bf n} relative to the system's center of mass.
Of course all these quantities must be calculated for either
sphere.

Each sensor output is affected by thermal and back-action displacement
noise spectral densities, which must be formed for both spheres:

\begin{eqnarray}
S_{uu}^{[th + ba]}(\omega) & = & \sum_{nl}\,\frac{2l+1}{4\pi M}\,
|A_{nl}(a)|^2\,|L_{nl}(\omega)|^2\,\left[\frac{2kT\omega_{nl}^2}{Q_{nl} \, \omega}\right.
\nonumber \\
& + & \left.\frac{2l+1}{4\pi M}\,|A_{nl}(a)|^2\,\sum_j\,
|{\cal P}_l({\bf n\!\cdot\! n}_j)|^2\,S_{FF}^{ba}\right],
\label{eq2}
\end{eqnarray}
%
%
where $k\/$ is Boltzmann's constant, $T\/$ the sphere's thermodynamic
temperature, and $M\/$ the sphere's mass, whether solid or hollow.
$S_{FF}^{ba}$ is the back action force spectral density, ${\cal P}_l$
a Legendre polynomial, and ${\bf n}_j$ the spherical coordinates of the
optical cavities ({\bf n}\,$\equiv$\,{\bf n$_1$}). The sum over $j\/$
accounts for the fact that each sensor is additionally affected by the
back-action noise forces exerted by the others. The back-action noise
force is given by

\begin{equation}
S_{FF}^{ba}(\omega) = \frac{4}{c^2 \, \pi^2}\,(1-\zeta)^2\,
h\nu_l\,F^2\,P\,
\frac{1}{1+\biggl(\frac{2 F L_c \omega}{\pi c}\biggr)^2} \ ,
\end{equation}
where $\nu_l$ is the light frequency, $c$ the speed of light, $P$
the light power incident on the cavity and $\zeta^2$ the fraction of light
reflected by the cavity at its resonance.

Assuming the noise in the spheres is uncorrelated~\cite{foot},
the total displacement spectral density is the sum of expressions
like~Eq.~(\ref{eq2}) for each sphere, plus a photodetector noise term,
the {\itshape shot noise}. The total {\itshape{strain}} noise spectral
density is thus given by

\begin{equation} 
S_{hh}(\omega) = \frac{S_{uu,{\rm hollow}}^{[th + ba]}(\omega) +
S_{uu,{\rm solid}}^{[th + ba]}(\omega) + S_{uu}^{shot}(\omega)}
{|u_{\rm hollow}(\omega)- u_{\rm solid}(\omega)|^2/|\tilde{h}(\omega)|^2} \ .
\label{eq5}
\end{equation} 
Here $S_{uu}^{shot} (\omega)$ is the photodetector noise which can be
written as

\begin{equation}
S_{uu}^{shot}(\omega) = 4\cdot 10^{-33} \biggl[
1+\biggl(\frac{2 F L_c \omega}{\pi c}\biggr)^2\biggr]\,
\frac{1}{F^2 P}\ {\rm \frac{m^2}{Hz}}   \ .
\end{equation}

\begin{figure}
\centerline{\psfig{figure=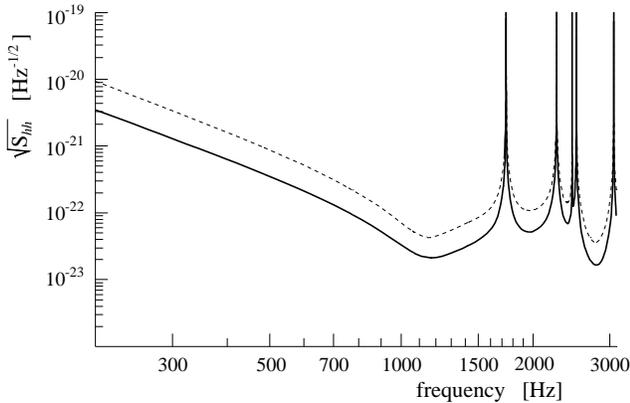,width=8.3cm}}
\caption{Spectral strain noise of a molybdenum dual sphere GW detector equipped with
$10^6$ finesse, 2~cm long Fabry-Perot cavities as motion sensors. Fundamental mode frequencies: 1100~Hz for the
outer (hollow) sphere, 2825~Hz for the solid (inner) sphere; total mass: $29+8$~tons. Solid curve: SQL imposed at 1.3~kHz, with $P\/$\,=\,7\,W, $Q/T \, \geq \, 1.6\cdot 10^8$\,K$^{-1}$.
Dashed curve: $P\/$\,=\,1\,W, $Q/T$~=~2$\cdot$10$^7$~K$^{-1}$.}
\label{fig2} 
\end{figure}

We may now consider the actual GW sensitivity of a system like this. We
take as reference an operation at the Standard Quantum Limit. The SQL is
reached at laser powers such that the shot noise and the radiation pressure
equally contribute to the total noise and, at the same time, the back-action
noise overcomes the mechanical resonator thermal noise. This is the procedure envisioned also for interferometric GW detectors~\cite{Hough}. Note that, since
the cavity is short, order of 1 $cm$, the finesse can be made very high,
order of $10^6$ and beyond, before loosing signal strength, and  thus the
SQL can be approached at laser powers of the order of 10\,W. Moreover, since
the bandwidth $\Delta f$ is expected to be wide, $\Delta f$\,$\simeq$\,$f$,
the SQL condition for the mechanical resonators $kT/Q\/$\,=\,$h\Delta f/4$
allows $Q/T$\,=\,10$^8$\,K$^{-1}$. Given
that one is able to approach the SQL in the readout, then one needs a
large cross-section to GWs. The resonance frequency fixed, the latter scales
as $\rho v_s^5$, where $\rho$ is the density of the material and $v_s$ the
velocity of sound. 

Molybdenum shows up as an interesting choice, with a sound velocity $v_s$~=~6.2~km/s and a density
$\rho$~=~10$\cdot$10$^3$~kg/m$^3$. The fabrication of the dual sphere may procede from Mo powders, which can be pressed, sintered to 95$\%$-density and hot formed to custom shapes. This procedure allows mechanical Q of 2$\cdot$10$^7$ at temperatures $\leq$4~K~\cite{duffy}, necessary to approach the SQL as above discussed. The thermal properties of molybdenum are such that no particular difficulties are expected to arise when cooling down such large masses~\cite{Moduffy}. Building on the experience developed for vibrationally insulating masses of few tons, as in bar detectors, we are confident that a suitable design can be made for suspending few tens of tons. 

Hot pressed, sintered beryllium is even more interesting, as $v_s$~=~13~km/s and $\rho$~=~1.9$\cdot$10$^3$~kg/m$^3$; it has already been used in large sizes (over 1.3~m) and its thermal properties make it viable as well. The low-temperature mechanical $Q$ is going to be investigated~\cite{fross}. Another interesting material
is sapphire ($v_s$\,=\,10\,km/s, $\rho$\,=\,$4 \cdot 10^3$\,kg/m$^3$) which is already known to show $Q\/$\,$>$\,10$^8$ at $T\/$\,$<$\,10\,K. Sapphire, acting as substrate
of the mirrors, would also minimize thermoelastic effects at low 
temperatures~\cite{Cerdonio}. Sapphire drawback mainly resides on the difficulty of growing large enough crystals and/or joining together several pieces while preserving the high mechanical Q. We note that the $\rho v_s^5$ factor for beryllium is a factor of 2 greater than for sapphire.
   
A molybdenum detector with $R$~=~0.95~m and $a$~=~0.57~m, with a
small gap in between to place the motion sensing optical cavities, would
give an interestingly low strain spectral noise in a rather wide frequency
band in the kHz region, see Fig.~\ref{fig2}. Here we plot the SQL spectral
strain noise, when the radiation pressure
noise is matched to the shot noise at 1.3~kHz: this requires an input light power of 7~W and $Q/T$~$\geq$~2$\cdot$10$^8$~K$^{-1}$. The spectral strain noise
is also shown for a
lower light power $P\/$\,=\,1\,W, $Q/T$~=~2$\cdot$10$^7$~K$^{-1}$ possibly more amenable to cryogenic
operation at $T\/$\,$\simeq$\,1\,K, but still giving an interesting
performance. Note that the spectral sensitivity is contaminated by the
thermal noise peaks of the non-quadrupole resonances.
The problem of unwanted narrow resonances in the sensitive frequency band
is also present in the case of interferometric detectors and methods
have been devised to filter them out~\cite{virgo,finn}.

\begin{figure} 
\centerline{\psfig{figure=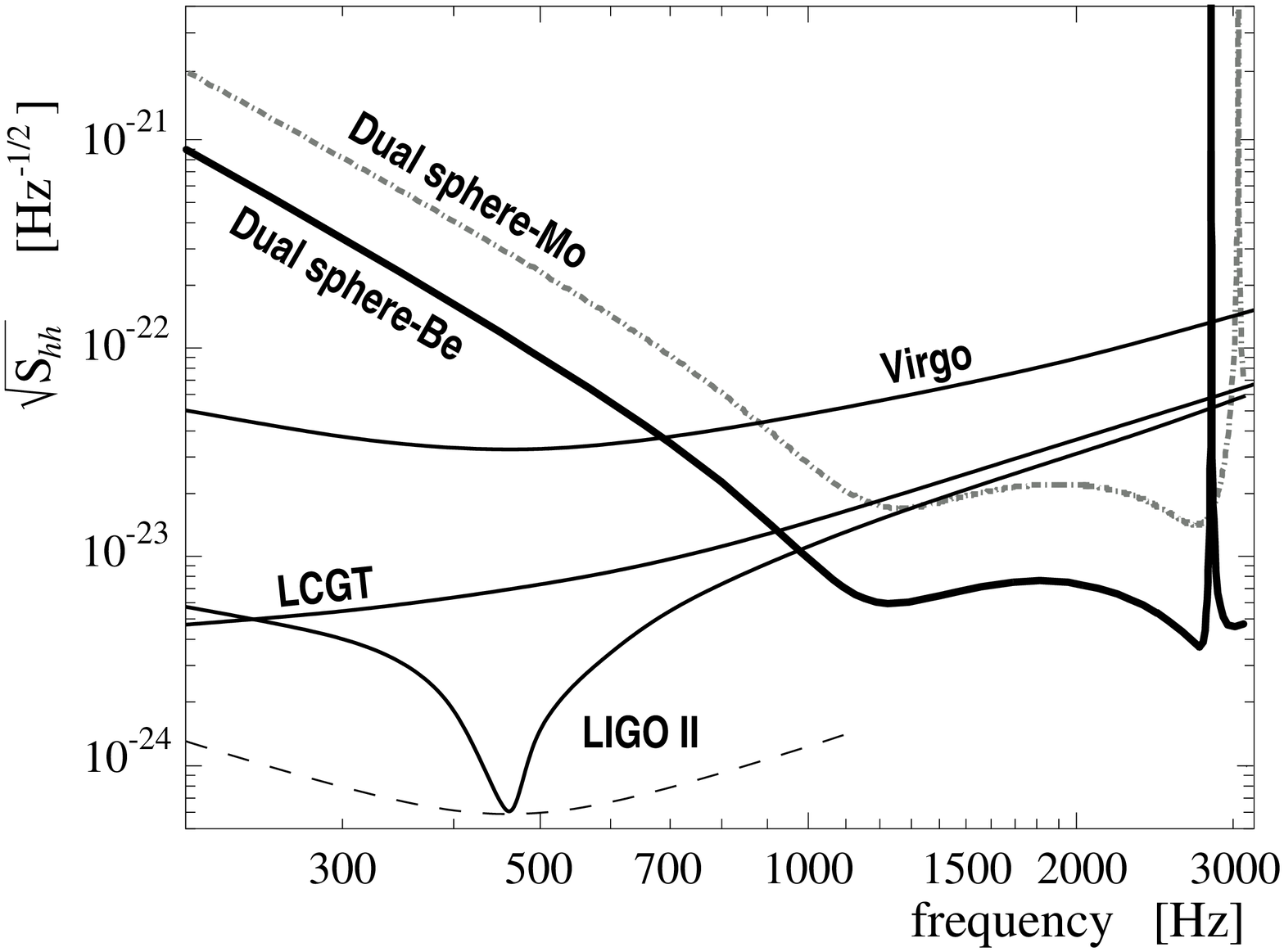,width=8.3cm}}
\caption{Spectral strain noise of GW detectors to come compared to dual spheres of molybdenum and beryllium. The LIGO~II curve corresponds to the dual-recycled operation tuned to 500~Hz. The dashed line is the location of the minimum for different LIGO~II tuning frequencies.}\label{fig3}
\end{figure}

Figure~\ref{fig3} shows a comparative plot of the sensitivities of
various GW detectors to come: initial VIRGO~\cite{virgo}, the cryogenic
interferometer LCGT~\cite{lcgt} and LIGO~II operated in the narrowband dual-recycled mode~\cite{ligo2}. For the proposed dual sphere we show the SQL operation of the SQL molybdenum system of Fig.~\ref{fig2} with the non-quadrupole resonances 
suppressed for clarity. The large drop in sensitivity,
indicated by the prominent spike towards the right end of the figure
is due to signal cancellation at a frequency $\omega_\star$ for which
$u_{\rm hollow}(\omega_\star)=u_{\rm solid}(\omega_\star)$, which causes the
denominator in eq.~(\ref{eq5}) to vanish. The presence of such a frequency
$\omega_\star$ is expected on the basis of the intrinsic structure of
eq.~(\ref{eq5}), and therefore it must be taken into consideration when one
chooses materials and dimensions. We also show in Fig.~\ref{fig3} the sensitivity of a beryllium system operated at SQL for 1300~Hz, which  requires an input light power of 12~W and $Q/T$~=~2$\cdot$10$^8$~K$^{-1}$; the gain in sensitivity is due to the larger $\rho v_s^5$ factor. As it can be seen, the dual sphere system
well compares with the best foreseen GW detectors, especially in the
high frequency region where e.g.\ relatively small mass $10M_\odot$
BH-BH mergers are expected~\cite{eana}.

In the end, the system we propose may still look like a two-mode system
in that the most useful band is obtained between the two first
quadrupole resonances of the two spheres. However the concept we propose
allows one to choose such two frequencies with a lot of freedom, and in
fact to open considerably the band with respect to systems which make use
of resonant secondary masses to get the two-mode operation.

We have noted that crucial practical issues for the realization
of such a massive system, especially in respect to fabrication, suspensions, cooling, can be dealt with; however the level of practicability of the concept may still strongly depend on supportive research.

We thank Stefano Vitale and Michele Bignotto for fruitful discussions. J.A.\ L. is grateful to INFN Legnaro National Laboratories for hospitality within the E.U programme "TMR - Access to LSF" (Contract ERBFMGECT980110). He also thanks the Spanish MEC for partial support, contract number PB96-0384.


\begin{references}
\bibitem{Req}  In the language and notation of ref.~\cite{Price}, $\Delta f/f=R_{eq}$ where $R_{eq}$ is the dimensionless noise resistance. For the readout systems of the past $R_{eq}\/$\,=\,10$^{-5}\div10^{-7}$ while for a 1~cm-long Fabry-Perot cavity 
with finesse $F\/$\,$\sim$\,10$^6$, input power $P\/$\,$\sim$\,10\ $W$, $\lambda = 1.064\ \mu$m it is possible to 
have $R_{eq}$ as large as $10^{-1}$.
\bibitem{foot} Actually, the back-action noise force acting on the cavity
mirrors is correlated because the cavity decay time is shorter or at most
comparable to the inverse of the frequencies of interest. In any case the
correction due to the non-null correlation is estimated to be small and is
neglected for simplicity.
\end{references}
\end{document}